\documentclass[11pt,preprint]{elsarticle}

\usepackage{fancyhdr}
\usepackage{fullpage}
\usepackage{amsmath}
\usepackage{amsbsy}
\usepackage{amssymb}
\usepackage{amscd}
\usepackage{amsfonts}
\usepackage{supertabular}
\usepackage{graphics}
\usepackage{verbatim}
\usepackage{xspace}
\usepackage{euscript}
\usepackage{alltt}
\usepackage{boxedminipage}
\usepackage{float}
\usepackage[colorlinks]{hyperref}
\usepackage{color}
\usepackage[all]{xy}
\usepackage{t1enc}
\usepackage{times,exscale}
\usepackage{graphicx,calc}
\usepackage{subfig}
\usepackage[ruled,vlined]{algorithm2e}

\def\bR{{\mathbf{R}}}
\def\bx{{\mathbf{x}}}
\def\R{{\mathbb{R}}}

\begin{document}

\begin{frontmatter}

\title{Augmented Lagrangian formulation of Orbital-Free Density Functional Theory}
\author[gatech]{Phanish Suryanarayana\corref{cor}}
\author[gatech]{Deepa Phanish}
\address[gatech]{College of Engineering, Georgia Institute of Technology, GA 30332, USA}
\cortext[cor]{Corresponding Author (\it phanish.suryanarayana@ce.gatech.edu) }

\begin{abstract}
We present an Augmented Lagrangian formulation and its real-space implementation for non-periodic orbital-free Density Functional Theory (OF-DFT) calculations. In particular, we rewrite the constrained minimization problem of OF-DFT as a sequence of minimization problems without any constraint, thereby making it amenable to powerful unconstrained optimization algorithms.  Further, we develop a parallel implementation of this approach for the Thomas-Fermi-von Weizscaker (TFW) kinetic energy functional in the framework of higher-order finite-differences and the conjugate gradient method. With this implementation, we establish that the Augmented Lagrangian approach is highly competitive compared to the penalty and Lagrange multiplier methods. Additionally, we show that higher-order finite-differences represent a computationally efficient discretization for performing OF-DFT simulations. Overall, we demonstrate that the proposed formulation and implementation is both efficient and robust by studying selected examples, including systems consisting of thousands of atoms. We validate the accuracy of the computed energies and forces by comparing them with those obtained by existing plane-wave methods. 
\end{abstract}


\end{frontmatter}

\section{Introduction}
Electronic structure calculations based on Kohn-Sham Density Functional Theory (DFT) \cite{Hohenberg, Kohn1965} have been remarkably successful in describing material properties and behavior. In DFT, the system of interacting electrons is replaced with a system of non-interacting electrons moving in an effective potential. As a result, the problem of determining a single many-electron wavefunction reduces to the calculation of multiple single-electron wavefunctions/orbitals, the number of which commensurates with size of the system \cite{Parr1989,NumAnalysis2003,Martin2004}. Solving for these orbitals using traditional approaches like diagonalization results in a cubic-scaling with respect to the number of atoms \cite{NumAnalysis2003,Martin2004}. To overcome this restrictive scaling, substantial effort has been directed towards the development of linear-scaling methods \cite{Goedecker, Bowler2012}. However, an efficient linear-scaling method for metallic systems still remains an open problem \cite{Cances2008}. Orbital-free DFT (OF-DFT) represents a simpler linear-scaling version of DFT wherein the electronic kinetic energy is modeled using a functional of the electron density \cite{Thomas1927,Fermi1927,Weizsacker1935,Teter1992,WangBook2002}. Detailed studies have shown that such a theory can provide an accurate description of systems whose electronic structure resembles a free-electron gas e.g. Aluminum \cite{carling2003orbital,ho2007energetics,huang2008transferable}. Current efforts include extending the applicability of OF-DFT to covalently bonded materials \cite{zhou2005improving} as well as molecular systems \cite{xia2012can}. 

OF-DFT calculations have traditionally employed the plane-wave basis \cite{Teter1992,Ho2008,Hung2010}, which is attractive because of the associated spectral convergence and the efficient evaluation of convolutions through the Fast Fourier Transform (FFT) \cite{Cooley1965}. However, the need for periodic boundary conditions limits the effectiveness of plane-waves in the study of isolated clusters and crystallographic defects. Further, the development of implementations which can efficiently utilize modern large-scale, distributed-memory computer architectures is particularly challenging. In view of this, recent efforts have been directed towards developing real-space approaches for OF-DFT, including finite-differences \cite{Garcia2007} and finite-elements \cite{Gavini2007,Motamarri2012}. Amongst these, the finite-element method provides the flexibility of an adaptive discretization. This attribute has been employed to perform all-electron calculations \cite{Gavini2007,Motamarri2012} and to develop a coarse-grained formulation of OF-DFT for studying crystal defects \cite{GaviniQC2007}. Notably, it has been demonstrated that higher-order finite-elements provide tremendous computational savings relative to their linear counterparts, especially for achieving chemical accuracies \cite{Motamarri2012}. However, higher-order finite-differences --- which have been particularly successful in DFT \cite{PARSEC,Alemany2004,OCTOPUS,suryanarayana2013spectral} due to their simplicity, efficiency and the ease with which the order of approximation can be changed --- remain unexplored in the context of OF-DFT.  

The electronic ground state in OF-DFT can be expressed as the solution of a non-linear, constrained minimization problem \cite{Garcia2007,Lieb1981,Blanc2005,Cances2012,Gavini2007}. Approaches that have been employed to solve this problem include steepest descents \cite{Ho2008}, conjugate-gradients \cite{Ho2008,Gavini2007,Jiang2004}, truncated Newton \cite{Ho2008,Garcia2007} and multigrid methods \cite{Ho2008}. Amongst these, the truncated Newton method has been found to be particularly effective, subject to the availability of a good starting guess \cite{Ho2008,Garcia2007}. In these aforementioned approaches, the techniques utilized to enforce the constraint include the penalty method \cite{Gavini2007}, the Lagrange multiplier method \cite{hung2012preconditioners,Motamarri2012} and its variant in terms of the projected gradient and Hessian \cite{Garcia2007}. In this work, we develop an Augmented Lagrangian \cite{Nocedal2006} formulation for OF-DFT which inherits the advantages of both the Lagrange multiplier and penalty methods. Specifically, we rewrite OF-DFT's constrained minimization problem as a sequence of unconstrained minimization problems which can be solved using efficient optimization algorithms like conjugate gradients. We develop a parallel implementation of the proposed method in the framework of  higher-order finite-differences. We demonstrate the robustness, efficiency and accuracy of the proposed approach through selected examples, the results of which are compared against the existing plane-wave methods. 

The remainder of this paper is organized as follows. We introduce OF-DFT in Section \ref{Sec:OFDFT} and discuss its solution  using the Augmented Lagrangian method in Section \ref{Sec:AugLag}. Subsequently, we describe the numerical implementation of the proposed method in Section \ref{Section:NumericalImplementation}, and validate it through examples in Section \ref{Section:Examples}. Finally, we conclude in Section \ref{Section:Conclusions}.

\section{Orbital-free Density Functional Theory} \label{Sec:OFDFT}
Consider a charge neutral system with $M$ atoms and $N$ electrons. Let $\bR = \{\bR_1, \bR_2, \ldots, \bR_M \}$ denote the positions of the nuclei with charges $\mathbf{Z} = \{Z_1, Z_2, \ldots, Z_M \}$ respectively. The energy of this system as described by OF-DFT is \cite{Parr1989}
\begin{equation} \label{Eqn:Energy:OFDFT}
\mathcal{E} (u, \bR) = T_s(u) + E_{\rm xc}(u) + E_{\rm H}(u) + E_{\rm ext}(u,\bR) + E_{\rm zz}(\bR) \,,
\end{equation}
where $u = \sqrt{\rho}$, $\rho$ being the electron density. The electronic kinetic energy $T_s(u)$ is typically modeled using the Thomas-Fermi-von Weizscaker (TFW) functional \cite{Parr1989} with or without additional terms that account for the linear-response of a uniform electron gas \cite{WangBook2002}. Examples include the Wang \& Teter (WT)\cite{Teter1992} and Wang, Govind \& Carter (WGC) \cite{Wang1998,Wang1999} kinetic energy functionals. For the non-periodic setting considered here, we restrict ourselves to the TFW family of functionals 
\begin{equation} \label{Eqn:KineticEnergy}
T_s(u) = C_F \int u^{10/3} (\bx) \, \mathrm{d\bx} + \frac{\lambda}{2} \int |\nabla u(\bx)|^2 \, \mathrm{d\bx} \,,
\end{equation} 
where $\lambda$ is an adjustable parameter and the constant $C_F = \frac{3}{10} (3\pi^2)^\frac{2}{3}$. 

The second term in Eqn. \ref{Eqn:Energy:OFDFT} is referred to as the exchange-correlation energy. In OF-DFT, it is generally modeled using the local density approximation (LDA) \cite{Kohn1965}:  
\begin{equation}
E_{xc} (u) = \int \varepsilon_{xc} (u(\bx)) u^2(\bx) \, \mathrm{d \bx} \,,
\end{equation}
where $\varepsilon_{xc} (u) = \varepsilon_x (u) + \varepsilon_c (u)$ is the sum of the exchange and correlation per particle of a uniform electron gas of density $\rho = u^2 $. Within the Ceperley-Alder parametrization \cite{Ceperley1980,PhysRevB.23.5048}, the exchange and correlation functionals have the following representation
\begin{eqnarray}
\varepsilon_{x}(u) & = & -\frac{3}{4}\left(\frac{3}{\pi}\right)^{1/3}u^{2/3} \,, \\
\varepsilon_{c}(u) & = & 
\begin{cases}
\frac{\gamma_1}{1+\beta_1\sqrt{r_s}+\beta_2{r_s}}\,\,\,\,\,\,\,\,\,\,\,\
r_s\geq 1\\
A_1\log{r_s}+B_1+C_1r_s\log{r_s}+D_1r_s\,\,\,\, r_s<1
\end{cases}
\end{eqnarray}
where $r_s=(\frac{3}{4\pi u^2})^{1/3}$, and the constants $\gamma_1=-0.1423$, ${\beta_1}=1.0529$, ${\beta_2}=0.3334$, $A_1=0.0311$, $B_1=-0.048$, $C_1=0.002$ and $D_1=-0.0116$.

The final three terms in Eqn. \ref{Eqn:Energy:OFDFT} represent electrostatic energies \cite{Martin2004}. In non-periodic systems, they can be expressed as 
\begin{eqnarray}
E_{\rm H}(u) & = & \frac{1}{2} \int \int \frac{u^2(\bx)u^2(\bx')}{|\bx - \bx'|} \,\mathrm{d\bx} \, \mathrm{d\bx'}\,, \label{Eqn:EH} \\
E_{\rm ext} (u,\bR) & = & \int u^2(\bx) \left( \sum_{J=1}^M V_{J}(\bx,\bR_J) \right) \, \mathrm{d\bx} \,, \label{Eqn:Eext}\\
E_{\rm zz}(\bR) & = & \frac{1}{2} \sum_{I=1}^{M} \sum_{\begin{subarray}{l} J=1 \\J \neq I \end{subarray}}^{M} \frac{Z_{I} Z_{J}}{|\bR_{I}-\bR_{J}|} \,, \label{Eqn:EZZ}
\end{eqnarray}
where the Hartree energy $E_{\rm H}(u)$ is the classical interaction energy of the electron density, $V_{J}(\bx,\bR_J)$ is the potential due to the nucleus positioned at $\bR_J$, $E_{\rm ext}(u,\bR)$ is the interaction energy between the electron density and the nuclei, and $E_{\rm zz}(\bR)$ is the repulsion energy between the nuclei. 

In all-electron calculations, the nucleus is treated as a point charge and therefore $V_{J}(\bx,\bR_J)$ is given by the Coulombic potential. However, the singular nature of this potential necessitates the use of a large number of basis functions for obtaining an accurate solution \cite{Gavini2007,Motamarri2012}. Additionally, the tightly bound core electrons are chemically inactive. Finally, the kinetic energy functionals in OF-DFT are not accurate for rapidly varying electron densities. Therefore, it is common to remove the core electrons and utilize an effective potential for $V_J(\bx,\bR_J)$, which is referred to as the pseudopotential \cite{Martin2004}. The absence of orbitals in OF-DFT requires that the pseudopotential be local, i.e. $V_J(\bx,\bR_J)$ depends only on the distance from the nucleus \cite{Ho2008}. In this work, we utilize the local pseudopotential approximation, wherein the charge density of the nuclei can be defined to be \cite{Pask2005,Phanish2010,Phanish2012}
\begin{equation} \label{Eqn:b:Pseudopotential}
\quad b(\bx,\bR) = \sum_{J=1}^M b_J (\bx,\bR_J) \,\,, \,\, \text{where} \,\, b_J(\bx,\bR_J) = \frac{-1}{4 \pi} \nabla^2 V_J (\bx,\bR_J) .
\end{equation}
Since the pseudopotential replicates the Coulombic potential outside the core cutoff radius $r_c$, $b_J(\bx,\bR_J)$ has compact support within a ball of radius $r_c$ centered at $\bR_J$ \cite{Pask2005}. 

The direct evaluation of the electrostatic terms in Eqns. \ref{Eqn:EH}, \ref{Eqn:Eext} and \ref{Eqn:EZZ} scales quadratically  with respect to the number of atoms. If the charge densities $b_J (\bx,\bR_J)$ do not overlap, a linear-scaling formulation can be developed by rewriting the total electrostatic energy as the following maximization problem \cite{Pask2005,Gavini2007,Phanish2010}
\begin{eqnarray}
E_{H}(u) + E_{\rm ext} (u,\bR) + E_{\rm zz}(\bR) & = & \sup_{\phi \in Y} \left\{ \frac{-1}{8 \pi} \int |\nabla \phi(\bx)|^2 \, \mathrm{d\bx} + \int(u^2(\bx)+ b(\bx,\bR)) \phi(\bx) \, \mathrm{d\bx} \right \} \nonumber \\
& - & \frac{1}{2}\sum_{J=1}^{M} \int b_J(\bx,\bR_J) V_J(\bx,\bR_J) \, \mathrm{d\bx} \,, \label{Eqn:ElectroststaticReformulation}
\end{eqnarray}
where $\phi(\bx)$ is referred to as the electrostatic potential, and $Y$ represents some appropriate space of functions. The last term in Eqn. \ref{Eqn:ElectroststaticReformulation} is required to negate the nonphysical self energy of the nuclei, i.e. the interaction energy of each nucleus with its own potential \cite{Gavini2007,Phanish2010,Pask2012}. The above reformulation requires non-overlapping charge densities in order for the repulsive energy to be correctly evaluated. If this condition is violated, we quantify the error incurred and nullify it by utilizing the procedure described in Appendix \ref{Appendix:Correct:RepulsiveEnergy}. With the above reformulation of the electrostatics, we arrive at \cite{Gavini2007,Phanish2010,Phanish2011}
\begin{equation} \label{Eqn:Variational:phi}
\mathcal{E}(u,\bR) = \sup_{\phi \in Y} \mathcal{F}(u,\bR,\phi) \,,
\end{equation}
where
\begin{eqnarray}
\mathcal{F}(u,\bR,\phi) & = & C_F \int u^{10/3} (\bx) \, \mathrm{d\bx} + \frac{\lambda}{2} \int |\nabla u(\bx)|^2 \, \mathrm{d\bx} + \int \varepsilon_{xc} (u(\bx)) u^2(\bx) \, \mathrm{d \bx} - \frac{1}{8 \pi} \int |\nabla \phi(\bx)|^2 \, \mathrm{d\bx}  \nonumber \\
& + & \int(u^2(\bx)+ b(\bx,\bR)) \phi(\bx) \, \mathrm{d\bx} - \frac{1}{2}\sum_{J=1}^{M} \int b_J(\bx,\bR_J) V_J(\bx,\bR_J) \, \mathrm{d\bx} .
\end{eqnarray}

The ground state of the system in OF-DFT is given by the variational problem \cite{Lieb1981, Gavini2007, Garcia2007, Cances2012}
\begin{eqnarray} \label{Eqn:VariationalProblem:OFDFT}
\mathcal{E}_0 =  \inf_{\bR \in \R^{3M}} \inf_{u \in \mathcal{X}} \mathcal{E}(u,\bR)  \,, \quad \mathcal{X} = \left \{u: u\in X,\, u \geq 0, \, \mathcal{C}(u)=0 \right \} \,,
\end{eqnarray} 
where $X$ is some appropriate space of functions and 
\begin{equation}
\mathcal{C}(u) = \int{u^2(\bx)} \, \mathrm{d\bx} - N
\end{equation} 
represents the constraint on the total number of electrons. The inequality constraint $u \geq 0$ is to ensure that $u$ is nodeless, i.e. does not possess a zero crossing. The mathematical analysis of this problem can be found in literature \cite{Lieb1981b,Blanc2005,Garcia2007,Gavini2007}.  The numerical solution of Eqn. \ref{Eqn:VariationalProblem:OFDFT} involves solving the electronic structure problem
\begin{equation} \label{Eqn:GroundStateElectronic}
\mathcal{E}^*(\bR) = \inf_{u \in \mathcal{X}} \mathcal{E}(u,\bR) \,,
\end{equation}
for every configuration of the nuclei encountered during the minimization 
\begin{equation} \label{Eqn:GroundStateSplit}
\mathcal{E}_0 = \inf_{\bR \in \R^{3M}} \mathcal{E}^*(\bR) .
\end{equation}
The forces on the nuclei while solving for their equilibrium configuration can be determined by using the relation \cite{Phanish2012}
\begin{eqnarray} 
{\bf f}_J & = & -\frac{\partial \mathcal{E}^*(\bR)}{\partial \bR_J} = -\frac{\partial \mathcal{F}(u^*,\bR,\phi^*)}{\partial \bR_J} \nonumber \\
 & = & - \int \frac{\partial b_J(\bx,\bR_J)}{\partial \bR_J}\left(\phi^*(\bx)-V_J(\bx,\bR_J)\right) \, \rm{d\bx} \nonumber \\
 & = & \int \nabla b_J(\bx,\bR_J) \left(\phi^*(\bx)-V_J(\bx,\bR_J)\right) \, \rm{d\bx} \,, \label{Eqn:Force:Nuclei}
\end{eqnarray}
where $u^*(\bx)$ is the minimizer of the variational problem in Eqn. \ref{Eqn:GroundStateElectronic}, $\phi^*(\bx)$ is the solution of the Poisson equation
\begin{equation}
\frac{-1}{4\pi} \nabla^2 \phi^*(\bx) = u^{*2}(\bx) + b(\bx,\bR) \,,
\end{equation}
and ${\bf f}_J$ denotes the force on the $J^{th}$ nucleus.  The last equality in Eqn. \ref{Eqn:Force:Nuclei} is obtained by utilizing the spherical symmetry of $b_J(\bx,\bR_J)$. Since $\nabla b_J(\bx,\bR_J)$ has compact support in a ball of radius $r_c$ centered at $\bR_J$, the calculation of the forces on all the nuclei also scales linearly with respect to the number of atoms. 


\section{Augmented Lagrangian formulation} \label{Sec:AugLag}
In this section, we focus on developing a solution methodology for the electronic structure problem given in Eqn. \ref{Eqn:GroundStateElectronic}, i.e. 
\begin{equation} \label{Eqn:VariationalProblem:FixedPositions1}
\inf_{u \in \mathcal{X}} \mathcal{E}(u,\bR) \,, \quad \mathcal{X} = \left \{u: u\in X,\, u \geq 0, \, \mathcal{C}(u)=0 \right \} .
\end{equation}
While solving this nonlinear constrained minimization problem, the approaches that have been previously employed to enforce the constraint $\mathcal{C}(u)=0$ include the penalty method \cite{Gavini2007}, the Lagrange multiplier method \cite{Ho2008,hung2012preconditioners,Motamarri2012} and its variant in terms of projected gradient and Hessian \cite{Garcia2007}. Here, we propose the use of the Augmented Lagrangian method outlined in Algorithm \ref{Algo:AugLag}, which inherits the advantageous features of both the penalty and Lagrange multiplier approaches. \\ 

\begin{algorithm}[H] \label{Algo:AugLag}
{\bf Input}: $\mu_0 > 0$, $\eta_0$ and $0 < \kappa < 1$  \\
$q=0$ \\
\Repeat(){$|\mathcal{E}(u_{q+1},\bR) - \mathcal{E}(u_{q},\bR)| < tol$}
{
$u_{q+1} = \arg \inf_{u \in X} \mathcal{L}(u,\eta_q,\mu_q)$ \\
$\eta_{q+1} = \eta_q - \frac{\mathcal{C}(u_{q+1})}{\mu_q}$ \\
$\mu_{q+1} = \kappa \mu_q$, $\,\,(0 < \kappa < 1)$ \\
$q = q + 1$ \\
}
{\bf Output}: $\eta^* = \eta_{q+1}$ and $u^*(\bx) = u_{q+1}(\bx)$ 
\caption{Augmented Lagrangian method for OF-DFT}
\end{algorithm} \vspace{2mm}

We rewrite the constrained minimization problem in Eqn. \ref{Eqn:VariationalProblem:FixedPositions1} as a sequence of unconstrained minimization problems of the form 
\begin{equation} \label{Eqn:AugLag:VariationalProblem}
\inf_{u \in X} \mathcal{L}(u,\eta_q,\mu_q) \,, 
\end{equation}
where 
\begin{equation} \label{Eqn:AugLag}
\mathcal{L}(u,\eta_q,\mu_q) = \mathcal{E}(u,\bR) - \eta_q \mathcal{C}(u) + \frac{1}{2 \mu_q} \mathcal{C}^2(u) 
\end{equation}
is referred to as the Augmented Lagrangian \cite{Nocedal2006}. We incorporate the constraint $u \geq 0$ directly into the minimization algorithm for $\mathcal{L}(u,\eta_q,\mu_q)$, details of which can be found in Section \ref{Section:NumericalImplementation}. The first two terms of $\mathcal{L}(u,\eta_q,\mu_q)$ constitute the Lagrangian used in Lagrange multiplier methods, with $\eta_q$ being an estimate for the Lagrange multiplier. It is updated in every iteration using the relation
\begin{equation}
\eta_{q+1} = \eta_q - \frac{\mathcal{C}(u_{q+1})}{\mu_q} .
\end{equation}
The last term in $\mathcal{L}(u,\eta_q,\mu_q)$ is quadratic in nature and penalizes the violation of the constraint $\mathcal{C}(u)$ using an inverse penalty parameter $\mu_q$. The penalty for violation of the constraint is increased with every iteration by scaling $\mu_q$ with $0 < \kappa < 1$. We choose the convergence of the total energy to within a prespecified tolerance $tol$ as the stopping criteria for the iteration, and denote the solution so obtained by $\eta^*$ and $u^*(\bx)$.

In each iteration of the Augmented Lagrangian method, the unconstrained minimization problem given by Eqn. \ref{Eqn:AugLag:VariationalProblem} needs to be solved. The necessary condition for the minimizer is given by the Euler-Lagrange equation: 
\begin{equation} \label{Eqn:EulerLagrange}
\frac{\delta \mathcal{L}(u,\eta_q,\mu_q)}{\delta u} \equiv -\frac{\lambda}{2} \nabla^2 u(\bx) + \bigg( V_K(\bx)+V_{xc}(\bx)+\phi(\bx) \bigg) u(\bx) + \left(- \eta_q + \frac{1}{\mu_q} \mathcal{C}(u) \right) u(\bx) = 0 \,,
\end{equation}  
where $\phi(\bx)$ is the solution of the Poisson equation
\begin{equation} \label{Eqn:PoissonEL}
-\frac{1}{4\pi} \nabla^{2}\phi(\bx) = u^2(\bx) + b(\bx,\bR) \,, 
\end{equation}
and 
\begin{equation}
V_K(\bx) = \frac{5}{3} C_F u^{4/3}(\bx). 
\end{equation} 
The exchange-correlation potential $V_{xc}(\bx)$ can be decomposed as \cite{Parr1989,Martin2004}
\begin{equation}
V_{xc}(\bx) = V_x(\bx) + V_c(\bx) \,,
\end{equation}
where $V_x(\bx)$ is the exchange potential and $V_c(\bx)$ is the correlation potential. Within the Ceperley-Alder parametrization \cite{Ceperley1980,PhysRevB.23.5048}, they have the following representations 
\begin{eqnarray}
V_{x}(\bx) & = & -\left(\frac{3}{\pi}\right)^{1/3}u^{2/3}(\bx) \,, \\
V_{c}(\bx) & = & 
\begin{cases}
\frac{\gamma_1 + \frac{7}{6}\gamma_1 \beta_1 \sqrt{r_s(\bx)} + \frac{4}{3} \gamma_1 \beta_2 r_s(\bx)}{(1+\beta_1\sqrt{r_s(\bx)}+\beta_2{r_s(\bx)})^2}\,, \,\,\,\,\,\,\,\,\,\,\,\
r_s(\bx) \geq 1\\
\left( A_1 + \frac{2}{3} C_1 r_s(\bx) \right) \log r_s(\bx) + \left(B_1 - \frac{1}{3}A_1 \right) + \frac{1}{3} (2D_1-C_1)r_s(\bx)\,, \,\,\,\,\,\, r_s(\bx)<1
\end{cases}
\end{eqnarray}
where $r_s(\bx) = \left( \frac{3}{4 \pi u^2(\bx)} \right)^{1/3}$, and the constants are as given in Section \ref{Sec:OFDFT}. The sufficient condition for the solution of the Euler-Lagrange equation, which we denote by $u_{q+1}(\bx)$, to be a minimizer of Eqn. \ref{Eqn:AugLag:VariationalProblem} is given by the positivity of the second variation, i.e. 
\begin{equation}
\frac{\delta^2 \mathcal{L}(u,\eta_q,\mu_q)}{\delta^2 u} \bigg|_{u=u_{q+1}} > 0 .
\end{equation} 
Above, we have described the Augmented Lagrangian formulation of OF-DFT in terms of the TFW kinetic energy functional.  However, the approach is not restricted by the choice of functional, and is therefore applicable to the WT \cite{Teter1992} and WGC \cite{Wang1998,Wang1999} kinetic energy functionals. Real-space, linear-scaling implementations of these functionals would require their reformulation in terms of Helmholtz equations \cite{Choly2002} or the associated local variational principle \cite{Radhakrishnan2010,Motamarri2012}.

The Augmented Lagrangian method inherits the positive attributes of both the penalty and Lagrange multiplier approaches. On the one hand, unlike the Lagrange multiplier method where a saddle point problem needs to be solved, the Augmented Lagrangian technique and the penalty method require the solution of minimization problems. Consequently, powerful unconstrained optimization algorithms like conjugate gradients can be utilized. On the other hand, unlike the penalty method, convergence can be achieved at relatively large values of the inverse penalty parameter $\mu_q$, thereby circumventing the problem of ill-conditioning. Overall, the Augmented Lagrangian method possesses the desirable features to be an efficient and robust technique for performing OF-DFT simulations. This is indeed validated by the examples and results presented in Section \ref{Section:Examples}. 


\section{Numerical Implementation} \label{Section:NumericalImplementation}
In this section, we describe the numerical implementation of the Augmented Lagrangian formulation for OF-DFT in the framework of higher-order finite-differences. We restrict our computations to a cuboidal domain $\Omega$ of sides $L_1$, $L_2$ and $L_3$. We discretize $\Omega$ with a uniform finite-difference grid such that $L_1=n_1 h$, $L_2=n_2h$ and $L_3=n_3h$, where $h$  is the grid spacing and $n_1$, $n_2$ and $n_3$ are natural numbers. We index the grid points by $(i,j,k)$, where $i=1,2,\ldots, n_1$, $j=1,2,\ldots, n_2$ and $k=1,2,\ldots, n_3$. We approximate the Laplacian of any function $f$ at the grid point $(i,j,k)$ using higher-order finite-differences \cite{LeVeque2007}
\begin{eqnarray}\label{Eqn:FD:Laplacian}
\nabla^2 f \big|^{(i,j,k)} \approx \sum_{p=0}^n w_p \bigg(f^{(i+p,j,k)} + f^{(i-p,j,k)} + f^{(i,j+p,k)} + f^{(i,j-p,k)} + f^{(i,j,k+p)} + f^{(i,j,k-p)} \bigg) \,,
\end{eqnarray}
where $f^{(i,j,k)}$ represents the value of the function $f$ at the grid point $(i,j,k)$. The weights $w_p$ are given by \cite{Boyd1991,mazziotti1999spectral,jordan2003spectral}
\begin{eqnarray}
w_0 & = & - \frac{1}{h^2} \sum_{q=1}^n \frac{1}{q^2} \,, \nonumber \\
w_p & = & \frac{2 (-1)^{p+1}}{h^2 p^2} \frac{(n!)^2}{(n-p)! (n+p)!} \,, \,\, p=1, 2, \ldots, n.
\end{eqnarray} 
This finite-difference formula represents a $2n$ order accurate approximation of the Laplacian, i.e. error is $~ \mathcal{O}(h^{2n+1})$. For spatial integrations, we assume that the function is constant in a cube of side $h$ around each grid point, i.e.
\begin{equation}
\int_{\Omega} f(\bx) \, \mathrm{d\bx} \approx h^3  \sum_{i=1}^{n_1} \sum_{j=1}^{n_2} \sum_{k=1}^{n_3} f^{(i,j,k)} .
\end{equation}

We generate the starting guess $u_0^{(i,j,k)} = \sqrt{\rho_0^{(i,j,k)}}$ by superimposing the isolated-atom electron densities for the given positions of the nuclei. We do so by visiting each atom and projecting the isolated atom electron density on to the finite-difference nodes lying within a ball of prespecified radius centered at that nucleus. Similarly, we calculate the charge density of the $J^{th}$ nucleus in a ball of radius $r_c$ centered at $\bR_J$ by utilizing the expression
\begin{equation}
b_{J}^{(i,j,k)} = -\frac{1}{4\pi} \sum_{p=0}^n w_p \bigg(V^{(i+p,j,k)}_{J} + V^{(i-p,j,k)}_{J} + V^{(i,j+p,k)}_{J} + V^{(i,j-p,k)}_{J} + V^{(i,j,k+p)}_{J} + V^{(i,j,k-p)}_{J} \bigg) .
\end{equation}
We then calculate the total charge density of the nuclei by summing over the individual densities, i.e.
\begin{equation}
b^{(i,j,k)} = \sum_{J=1}^M b_{J}^{(i,j,k)} .
\end{equation}
The localized nature of the above calculations ensure that the evaluation of $u_0^{(i,j,k)}$ and $b^{(i,j,k)}$ scales linearly with the number of atoms. 

In each iteration of the Augmented Lagrangian method, we discretize the resulting Euler-Lagrange equation (Eqn. \ref{Eqn:EulerLagrange}) as follows
\begin{eqnarray} \label{Eqn:EulerLagrangeDiscretized}
& & -\frac{\lambda}{2}\sum_{p=0}^n w_p \bigg(u^{(i+p,j,k)} + u^{(i-p,j,k)} + u^{(i,j+p,k)} + u^{(i,j-p,k)} + u^{(i,j,k+p)} + u^{(i,j,k-p)} \bigg) \nonumber \\
& & + \bigg( V_K^{(i,j,k)}+V_{xc}^{(i,j,k)}+\phi^{(i,j,k)} \bigg) u^{(i,j,k)} + \left(- \eta_q + \frac{1}{\mu_q} \mathcal{C}_h(u) \right) u^{(i,j,k)} = 0 \,,
\end{eqnarray}
where 
\begin{equation}
\mathcal{C}_h(u) = h^3 \sum_{i=1}^{n_1} \sum_{j=1}^{n_2} \sum_{k=1}^{n_3} (u^{(i,j,k)})^2 - N
\end{equation}
is the discretized constraint on the number of electrons. We employ zero Dirichlet boundary conditions for $u(\bx)$, owing to its rapid decay. This translates to setting $u^{(i,j,k)}=0$ for any index which does not correspond to a node in the finite-difference grid.  We solve the resulting set of equations using the Polak-Ribiere variant of non-linear conjugate gradients with a secant line search \cite{Shewchuk1994} and a Jacobi preconditioner \cite{Axelsson1985}. We maintain $u(\bx) \geq 0$ by replacing $u^{(i,j,k)}$ with $|u^{(i,j,k)}|$ after every update in the conjugate gradient method \cite{Garcia2007,Ho2008}. The solution of Eqn. \ref{Eqn:EulerLagrangeDiscretized} so obtained ($u^{(i,j,k)}_{q+1}$) is used as a starting guess for the next iteration within the Augmented Lagrangian method. 

We evaluate the electrostatic potential $\phi^{(i,j,k)}$ for every update of $u^{(i,j,k)}$ in the conjugate gradient method, an approach which has been shown to be more robust and stable compared to the simultaneous solution of $u^{(i,j,k)}$ and $\phi^{(i,j,k)}$ \cite{Motamarri2012}. We discretize the Poisson equation (Eqn. \ref{Eqn:PoissonEL}) on the finite-difference grid as
\begin{equation} \label{Eqn:Poisson:Discretized}
- \frac{1}{4 \pi} \sum_{p=0}^n w_p \bigg(\phi^{(i+p,j,k)} + \phi^{(i-p,j,k)} + \phi^{(i,j+p,k)} + \phi^{(i,j-p,k)} + \phi^{(i,j,k+p)} + \phi^{(i,j,k-p)} \bigg) = (u^{(i,j,k)})^2 + b^{(i,j,k)} .
\end{equation}
We utilize zero Dirichlet boundary conditions for the electrostatic potential, which translates to setting $\phi^{(i,j,k)}=0$ for any index which does not correspond to a node in the finite-difference grid. We solve the resulting linear system of equations using the Generalized minimal residual method (GMRES) \cite{saad1986gmres}.  For every subsequent Poisson equation encountered, we use the previous solution as the starting guess. 

After determining the electronic ground state, i.e. $u^{*(i,j,k)}$ and $\phi^{*(i,j,k)}$, we evaluate the forces on the nuclei using Eqn. \ref{Eqn:Force:Nuclei}. In order to do so, we calculate $\nabla b_J(\bx,\bR_J)$ in a ball of radius $r_c$ centered at $\bR_J$ by utilizing a higher order finite-difference approximation:
\begin{eqnarray}
\nabla b_J \big|^{(i,j,k)} \approx \sum_{p=1}^n \tilde{w}_p \bigg( ( b_J^{(i+p,j,k)} - b_J^{(i-p,j,k)}) \hat{\mathbf{e}}_1 + ( b_J^{(i,j+p,k)} - b_J^{(i,j-p,k)}) \hat{\mathbf{e}}_2  + ( b_J^{(i,j,k+p)} - b_J^{(i,j,k-p)}) \hat{\mathbf{e}}_3 \bigg) \,,
\end{eqnarray}
where $\hat{\mathbf{e}}_1$, $\hat{\mathbf{e}}_2$ and $\hat{\mathbf{e}}_3$ represent unit vectors along the edges of the cuboidal domain $\Omega$. The weights $\tilde{w}_p$ are given by \cite{Boyd1991,mazziotti1999spectral,jordan2003spectral}
\begin{equation}
\tilde{w}_p = \frac{(-1)^{p+1}}{h p} \frac{(n!)^2}{(n-p)! (n+p)!} \,, \,\, p=1, 2, \ldots, n.
\end{equation}
This finite-difference expression represents a $2n$ order accurate approximation of the gradient operator, i.e. error is $~ \mathcal{O}(h^{2n+1})$. We determine the ground state configuration of the nuclei by using the Polak-Ribiere variant of non-linear conjugate gradients with a secant line search \cite{Shewchuk1994}.  

We have developed a parallel implementation of the proposed approach using the Portable, Extensible Toolkit for scientific computations (PETSc) \cite{Petsc1,Petsc2} suite of data structures and routines. Within PETSc, we have utilized distributed arrays with the star-type stencil option. The communication between the processors is handled via the Message Passing Interface (MPI). 

\section{Examples and Results} \label{Section:Examples}
In this section, we use selected examples to validate the proposed Augmented Lagrangian formulation and higher-order finite-difference implementation of OF-DFT. For all the simulations, we choose the TFW kinetic energy functional with $\lambda=0.2$, which has been found to be the most appropriate value for isolated systems \cite{Parr1989}. Further, we employ the pseudopotential approximation through the Goodwin-Needs-Heine pseudopotential \cite{Goodwin1990}. In the discussions below, we refer to the implementation developed in this work as RS-FD, which is an acronym for Real-Space Finite-Differences.

\subsection{Convergence with domain size} \label{Subsec:ConvergenceDomain}
The theory of OF-DFT as presented in Section \ref{Sec:OFDFT} is for all of space, i.e. $\R^3$. Since practical calculations are restricted to finite regions, it is necessary to verify the convergence of the results with respect to the size of the computational domain $\Omega$. For the examples considered in this work, we choose $5 \times 5 \times 5$ FCC unit cells of Aluminum with lattice constant $a = 8.0$ Bohr as the representative system. In Table \ref{Table:ConvergenceDomain}, we demonstrate that there is rapid convergence in the energy with domain size. In fact, the error due to the finite domain is less than $0.0001$ eV/atom when every atom is at least $12.0$ Bohr from the boundary. In view of this, for all the remaining simulations presented in this paper, we choose $\Omega$ such that the minimum distance of any atom to the boundary is $12.0$ Bohr. 

\begin{table}[H]
\centering
\begin{tabular}{|c|c|c|c|c|c|c|}
\hline 
 $L$ (Bohr) & $56.0$ & $58.0$ & $60.0$ & $62.0$ & $64.0$ & $66.0$ \\
\hline 
 $\mathcal{E}$ (eV/atom) & $-59.9504$ & $-59.9458$ & $-59.9659$ & $-59.9627$ & $-59.9627$ & $-59.9627$ \\
\hline
\end{tabular}
\caption{Energy as a function of domain size for $5 \times 5 \times 5$ FCC unit cells of Aluminum with lattice constant $a=8.0$ Bohr. The domain $\Omega$ is a cube with sides $L_1=L_2=L_3=L$, the finite-difference approximation is sixth order accurate and mesh size $h=0.5$ Bohr.}
\label{Table:ConvergenceDomain}
\end{table}  

The rapid convergence of the energy with respect to domain size, even though zero Dirichlet boundary conditions have been imposed on the electrostatic potential $\phi(\bx)$, could be a consequence of the highly symmetric nature of the chosen cluster. Systems which possess a lower degree of symmetry, and in particular those that are polarized, may necessitate larger domain sizes within the current implementation. However, this requirement can be overcome by utilizing the multipole expansion to determine more accurate Dirichlet boundary conditions for $\phi(\bx)$ \cite{Burdick2003}. Another aspect worth noting is that the efficiency of the calculations can be significantly improved by imposing Dirichlet boundary conditions for $u(\bx)$ and $\phi(\bx)$ on an ellipsoid within the cuboidal domain $\Omega$.  In the above example, this translates to restricting the simulation to a sphere of diameter $L$. This would enable nearly a factor of two reduction in the number of finite-difference nodes, and consequently the computational time. 

\subsection{Convergence with spatial discretization} \label{Subsec:ConvergenceMesh}
In the finite-difference method, the order of the approximation determines the rate of convergence with respect to spatial discretization. Increasing the order typically results in a higher convergence rate. However, this comes at the price of increased computational cost per iteration due to the reduced locality of the discretized operators, which in the case of OF-DFT is the Laplacian. Here, we determine the order of the finite-difference approximation that will enable efficient OF-DFT calculations for the targeted accuracy of $0.005$ eV/atom in the energy. For this study, we again choose $5 \times 5 \times 5$ FCC unit cells of Aluminum with $a=8.0$ Bohr as the representative example. 

\begin{figure}[htbp]
\centering
\subfloat[Error in energy vs. mesh size]{\label{fig:ConvergenceMesh}\includegraphics[keepaspectratio=true,width=0.45\textwidth]{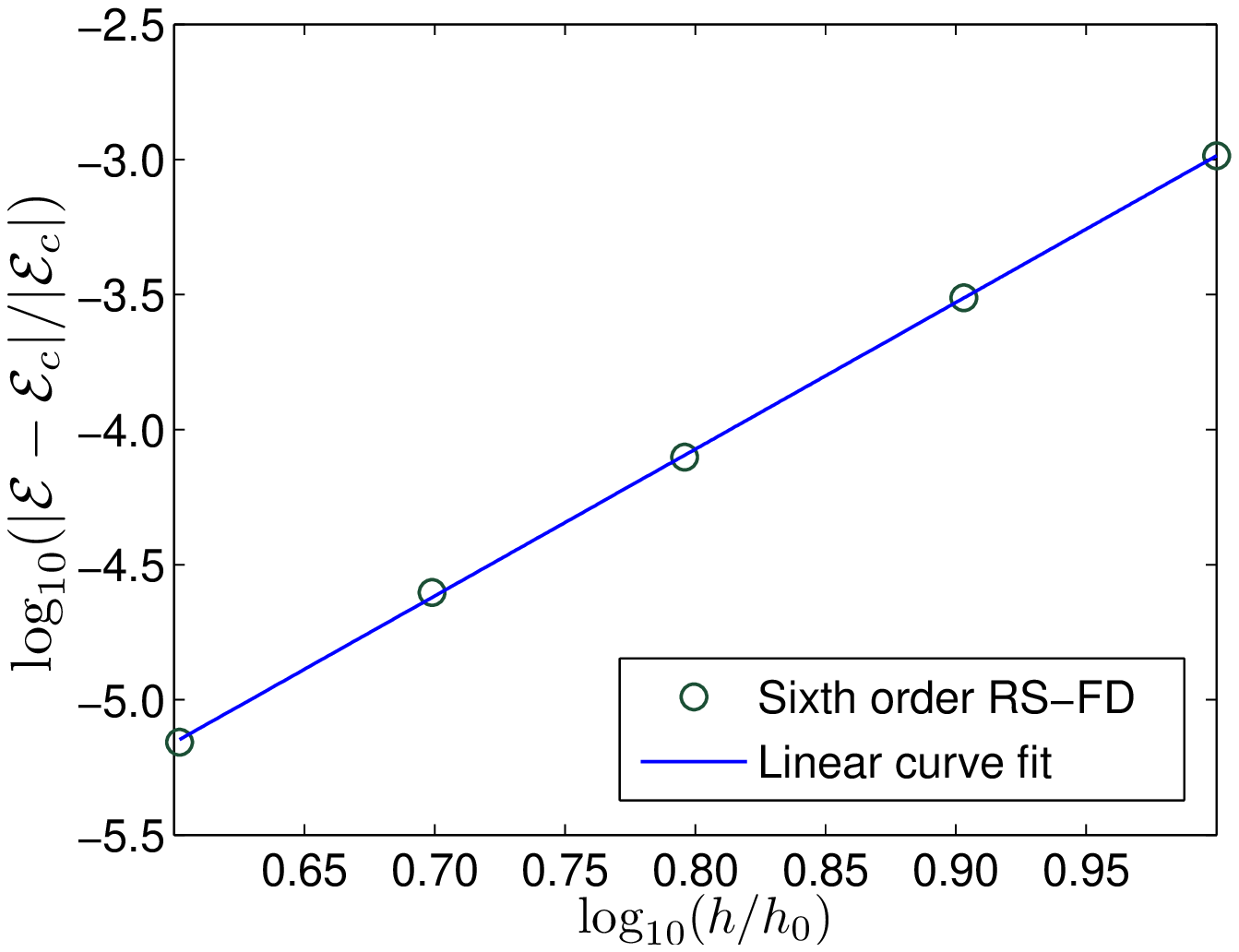}}
\subfloat[Error in energy vs. computational time]{\label{fig:ConvergenceTime}\includegraphics[keepaspectratio=true,width=0.45\textwidth]{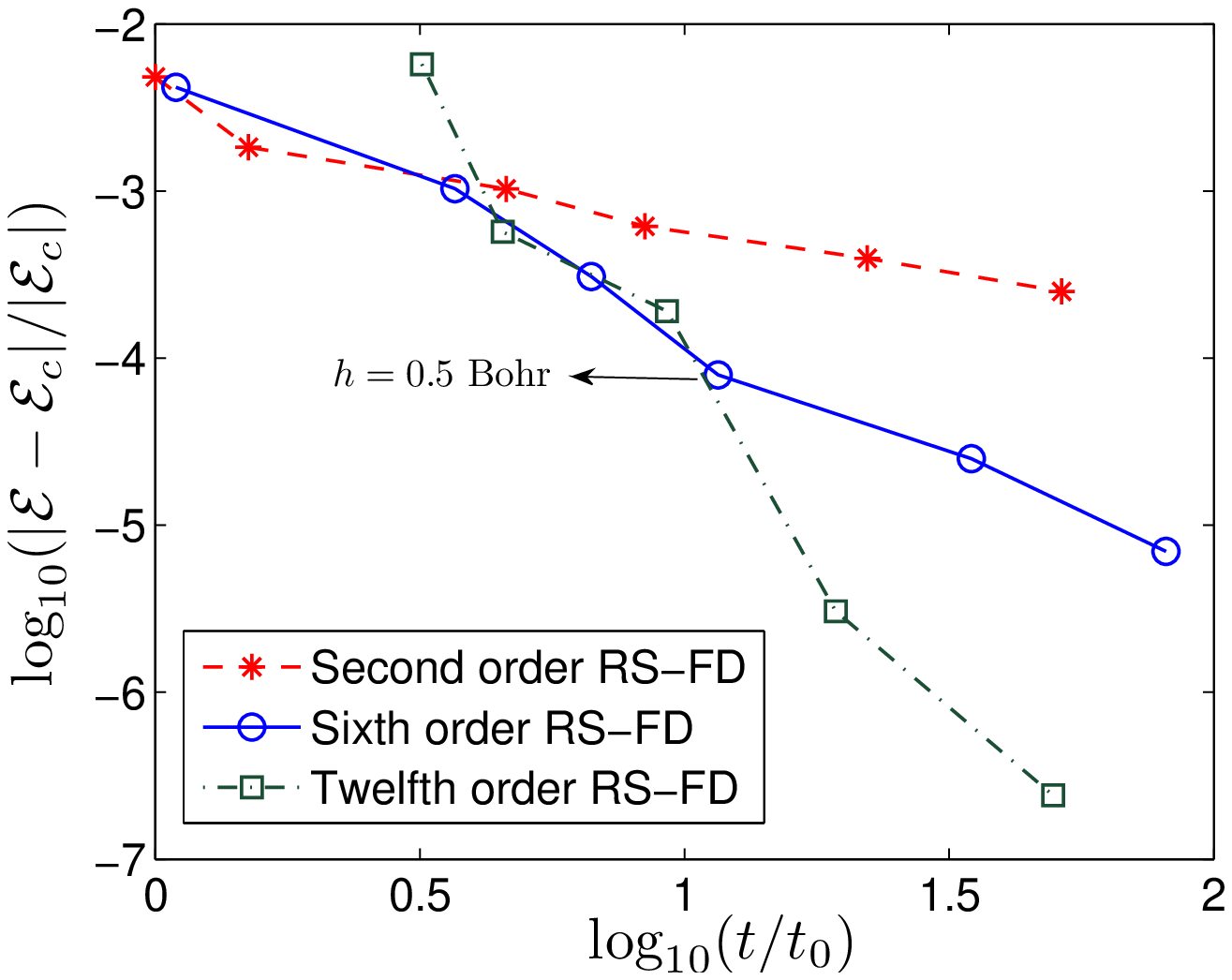}} \\
\caption{Convergence in the energy for different orders of finite-difference approximation. The system under consideration is $5 \times 5 \times 5$ FCC unit cells of Aluminum with lattice constant $a=8.0$ Bohr. }
\label{fig:ConvergenceMeshTime}
\end{figure}

We start by evaluating the energy of the aforementioned system for different mesh sizes $h$ while utilizing sixth order accurate finite-differences. Anticipating polynomial convergence with respect to mesh size, we fit the data to a power law of the form
\begin{equation}
\mathcal{E} = C h^p + \mathcal{E}_c, 
\end{equation} 
where the prefactor $C$, convergence rate $p$, and estimate of the $h \to 0$ energy $\mathcal{E}_c$ are the parameters to be fitted. For the example considered here, we obtain $C = -0.20$, $p=5.44$ and $\mathcal{E}_c=-59.95796$ eV/atom. Utilizing $\mathcal{E}_c$ as the reference value, we plot in Fig. \ref{fig:ConvergenceMesh} the convergence in energy with respect to discretization, where $h_0 = 0.08$ Bohr has been utilized to normalize the mesh size $h$. Next, we plot in Fig. \ref{fig:ConvergenceTime} the error in energy as a function of computational time for different orders of the finite-difference approximation. For the targeted accuracy of $0.005$ eV/atom, which corresponds to a relative accuracy of approximately $1 \times 10^{-4}$, sixth and twelfth order finite-differences have nearly identical performance. Notably, both are an order of magnitude more efficient than second order finite-differences. Since twelfth order finite-differences require larger inter-processor communication, we utilize sixth order finite-differences with $h=0.5$ Bohr in the remainder of this work. However, for situations that demand substantially higher accuracies, we expect twelfth order finite-differences to be the preferred choice. Overall, we conclude that the relatively high rates of convergence that can be achieved by higher-order finite-differences make them an attractive choice for performing electronic structure calculations based on OF-DFT. 

\subsection{Performance of the Augmented Lagrangian method}
We now analyze the capacity of the Augmented Lagrangian method to perform accurate, efficient and robust OF-DFT simulations. Specifically, we determine the sensitivity of the Augmented Lagrangian approach to choice of the input parameters in Section \ref{Subsubsec:ALParameters} and compare its performance to alternate solution strategies in Section \ref{Subsecsec:Comparison}. 

\subsubsection{Choice of input parameters} \label{Subsubsec:ALParameters}
As outlined in Algorithm \ref{Algo:AugLag}, the input parameters to the Augmented Lagrangian method are an initial estimate of the Lagrange multiplier $\eta_0$, inverse penalty parameter $\mu_0$ and scaling factor $\kappa$. Here, we analyze the effect of $\eta_0$, $\mu_0$ and $\kappa$ on the performance of the proposed method by comparing the number of matrix-vector products required to achieve a tolerance $tol = 0.0001$ eV/atom within the Augmented Lagrangian iteration. Since the total computational effort is dominated by matrix-vector products, it provides an accurate estimate of the relative execution time. In Tables \ref{Table:AugLagEta}, \ref{Table:AugLagMu} and \ref{Table:AugLagKappa}, we present the results so obtained for $5 \times 5 \times 5$ FCC unit cells of Aluminum with lattice constant $a=8.0$ Bohr. We have utilized sixth order accurate finite-differences with $h=0.5$ Bohr. It is clear from Table \ref{Table:AugLagEta} that as $\eta_0$ gets closer to the converged value $\eta^* = -0.1226$ Hartree, there is a reduction in the number of matrix-vector products to achieve convergence. Additionally, as demonstrated by the results in Table \ref{Table:AugLagMu}, choosing too small a value of $\mu_0$ can have a detrimental effect on the performance of the Augmented Lagrangian method. Finally, the number of matrix-vector multiplications is relatively independent of the choice of the scaling factor $\kappa$, as displayed by the results in Table \ref{Table:AugLagKappa}. Overall, the Augmented Lagrangian method produces identical results for various choices of  $\eta_0$, $\mu_0$ and $\kappa$. Therefore, it represents a robust approach for performing OF-DFT calculations. Henceforth, we choose $\eta_0=-0.2$, $\mu_0=1$ and $\kappa=0.1$, since this combination of parameters is found to perform appreciably.

\begin{table}[H]
\centering
\begin{tabular}{ccccc}
\hline 
 $\eta_0$ & $\mu_0$ & $\kappa$ & $\mathcal{E}$ & Matrix-vector  \\
(Hartree) & & & (eV/atom) & products \\ 
\hline
$-10.0$ & $1.0$ & $0.1$ & $-59.9627$ & $193848$ \\
$-1.0$ & $1.0$ & $0.1$ & $-59.9627$ & $134148$ \\
$-0.2$ & $1.0$ & $0.1$ & $-59.9627$ & $102162$ \\
$1.0$ & $1.0$ & $0.1$ & $-59.9627$ & $139740$ \\
$10.0$ & $1.0$ & $0.1$ & $-59.9627$ & $259884$ \\
\hline
\end{tabular}
\caption{Performance of Augmented Lagrangian method for different choices of $\eta_0$. The system under consideration is $5 \times 5 \times 5$ FCC unit cells of Aluminum with lattice constant $a=8.0$ Bohr. Sixth order accurate finite-differences with mesh size $h=0.5$ Bohr have been utilized.  }
\label{Table:AugLagEta}
\end{table}  

\begin{table}[H]
\centering
\begin{tabular}{ccccc}
\hline 
 $\eta_0$ & $\mu_0$ & $\kappa$ & $\mathcal{E}$ & Matrix-vector  \\
(Hartree) & & & (eV/atom) & products \\ 
\hline
$-0.2$ & $100$ & $0.1$ & $-59.9627$ & $115836$ \\
$-0.2$ & $10$ & $0.1$ & $-59.9627$ & $108773$ \\
$-0.2$ & $1$ & $0.1$ & $-59.9627$ & $102162$ \\
$-0.2$ & $0.1$ & $0.1$ & $-59.9627$ & $121332$ \\
$-0.2$ & $0.01$ & $0.1$ & $-59.9627$ & $252744$ \\
\hline
\end{tabular}
\caption{Performance of Augmented Lagrangian method for different choices of $\mu_0$. The system under consideration is $5 \times 5 \times 5$ FCC unit cells of Aluminum with lattice constant $a=8.0$ Bohr. Sixth order accurate finite-differences with mesh size $h=0.5$ Bohr have been utilized.  }
\label{Table:AugLagMu}
\end{table}  

\begin{table}[H]
\centering
\begin{tabular}{ccccc}
\hline 
 $\eta_0$ & $\mu_0$ & $\kappa$ & $\mathcal{E}$ & Matrix-vector  \\
(Hartree) & & & (eV/atom) & products \\ 
\hline
$-0.2$ & $1$ & $0.1$ & $-59.9627$ & $102162$ \\
$-0.2$ & $1$ & $0.2$ & $-59.9627$ & $102969$ \\
$-0.2$ & $1$ & $0.3$ & $-59.9627$ & $108652$ \\
$-0.2$ & $1$ & $0.4$ & $-59.9627$ & $102128$ \\
$-0.2$ & $1$ & $0.5$ & $-59.9627$ & $105548$ \\
\hline
\end{tabular}
\caption{Performance of Augmented Lagrangian method for different choices of $\kappa$. The system under consideration is $5 \times 5 \times 5$ FCC unit cells of Aluminum with lattice constant $a=8.0$ Bohr. Sixth order accurate finite-differences with mesh size $h=0.5$ Bohr have been utilized. }
\label{Table:AugLagKappa}
\end{table}

\subsubsection{Comparison with the penalty and Lagrange multiplier methods} \label{Subsecsec:Comparison}
As discussed in Section \ref{Sec:AugLag}, the Augmented Lagrangian approach combines the desirable characteristics of both the penalty and Lagrange multiplier methods. In view of this, a comparison of the performance of these approaches is merited in the context of OF-DFT. For this study, we implement the penalty method by setting $\eta_q=0$ throughout the Augmented Lagrangian iteration. Further, we implement the Lagrange multiplier based conjugate-gradient approach which was proposed for DFT \cite{payne1992iterative} and later adapted for OF-DFT \cite{Teter1992,jiang2004conjugate,Ho2008}. In particular, we choose the Polak-Ribiere variant of non-linear conjugate gradients \cite{Shewchuk1994} with Brent's method \cite{press2007numerical} for the line search. In Table \ref{Table:ComparisonApproaches}, we compare the number of matrix-vector products required to achieve convergence of $tol=0.0001$ eV/atom for three clusters, namely $1 \times 1 \times 1$, $3 \times 3 \times 3$ and $5 \times 5 \times 5$ FCC unit cells of Aluminum with lattice constant $a=8.0$ Bohr. For these examples, the Augmented Lagrangian approach demonstrates the best performance within our current implementations. We therefore conclude that the proposed Augmented Lagrangian method represents an efficient and accurate approach for performing OF-DFT calculations. 

In this work, we have solved the unconstrained minimization problem appearing in each iteration of the Augmented Lagrangian method using conjugate gradients. However, alternate optimization algorithms can be utilized, including the quadratically convergent Newton's method and its variants \cite{dennis1977quasi}. Since Newton's method requires a good starting guess, we envision a scheme where the conjugate gradient method is used in conjunction with Newton's method to be an attractive one. Another aspect worth noting is that the majority of the computational effort in the current implementation is devoted towards the repeated solution of the Poisson equation. When utilizing a Newton type method, it is possible to simultaneously solve for $u(\bx)$ and $\phi(\bx)$, which is expected to further improve the performance of the proposed formulation. However, such a technique has been shown to be less robust than the staggered approach adopted here \cite{Motamarri2012}. 

\begin{table}[H]
\centering
\begin{tabular}{cccc}
\hline 
 Cluster & Approach & $\mathcal{E}$ & Matrix-vector  \\
         &          &  (eV/atom)    &  products \\ 
\hline
                                      & Augmented Lagrangian Method      & $-59.2282$ & $16666$ \\
$1 \times 1 \times 1$ FCC unit cells  & Penalty Method                   & $-59.2282$ & $18705$ \\
                                      & Lagrange Multiplier Method       & $-59.2281$ & $18276$ \\
\hline
                                      & Augmented Lagrangian Method      & $-59.8085$ & $48285$ \\
$3 \times 3 \times 3$ FCC unit cells  & Penalty Method                   & $-59.8086$ & $55010$ \\
                                      & Lagrange Multiplier Method       & $-59.8085$ & $50838$ \\
\hline
                                      & Augmented Lagrangian Method      & $-59.9627$ & $102162$ \\
$5 \times 5 \times 5$ FCC unit cells  & Penalty Method                   & $-59.9627$ & $123456$ \\
                                      & Lagrange Multiplier Method       & $-59.9627$ & $129681$ \\
\hline
\end{tabular}
\caption{Comparison of the Augmented Lagrangian, penalty and Lagrange multiplier \cite{payne1992iterative} methods . Sixth order accurate finite-differences with mesh size $h=0.5$ Bohr have been utilized.}
\label{Table:ComparisonApproaches}
\end{table}  

\subsection{Examples: Aluminum clusters} 
In this section, we verify the accuracy of RS-FD by comparing its predictions with the plane-wave code PROFESS. Within RS-FD, we opt for sixth order accurate finite-differences with mesh size $h=0.5$ Bohr. Further, we choose the domain $\Omega$ such that the minimum distance of any atom to the boundary is $12$ Bohr. With this choice of parameters, the total energies and forces are converged to within $0.005$ eV/atom and $0.005$ eV/Bohr respectively. The binding energies, which we calculate with respect to the isolated atom energy computed with the same parameters, are converged to within $0.001$ eV/atom. Within PROFESS, we utilize a plane-wave energy cutoff $E_{cut}=600$ eV and the size of the domain is chosen such that the minimum distance of any atom to the boundary is $10$ Angstrom. The energies and forces so obtained are converged to within $0.0005$ eV/atom and $0.0005$ eV/Bohr respectively. 

We start with the relatively simple examples of $Al_2$ and $Al_3$ clusters. In Table \ref{Table:Al2_force}, we compare the forces obtained by RS-FD and PROFESS for different distances between the atoms in the $Al_2$ cluster.  There is very good agreement in the forces with the maximum difference being $0.001$ eV/Bohr. These results indicate that forces can be accurately calculated within our formulation and implementation. Next, we compare the equilibrium bond length and binding energy for the $Al_2$ and $Al_3$ clusters in Table \ref{Table:Al:Clusters}. The atoms in the $Al_3$ cluster are constrained to move such that the $D_{3h}$ symmetry group is maintained. Again, the results obtained by RS-FD are in excellent agreement with PROFESS. In fact, the equilibrium bond lengths and binding energies are identical to within $0.001$ Bohr and $0.001$ eV/atom respectively. 

\begin{table}[htbp]
\centering
\begin{tabular}{ccc}
\hline 
$R$ (Bohr) & $f$ (eV/Bohr) & $f$ (eV/Bohr)  \\
           &  RS-FD        & PROFESS  \\
\hline
$5.00$ & $-0.090$ & $-0.089$ \\
$5.08$ & $-0.043$ & $-0.043$ \\
$5.16$ & $+0.001$ & $+0.000$ \\
$5.24$ & $+0.040$ & $+0.041$ \\
$5.32$ & $+0.075$ & $+0.076$ \\    
\hline
\end{tabular}
\caption{Force ($f$) between the atoms in the $Al_2$ cluster for different interatomic distances ($R$). A negative force indicates repulsion whereas a positive one signifies attraction.}
\label{Table:Al2_force}
\end{table}

\begin{table}[htbp]
\centering
\begin{tabular}{cccccc}
\hline 
Cluster & $G$ & $\mathcal{E}_b$ (eV/atom) & $\mathcal{E}_b$ (eV/atom) & $R_e$ (Bohr) & $R_e$ (Bohr)  \\
        &     & RS-FD & PROFESS & RS-FD & PROFESS \\
\hline
$Al_2$ & $D_{\infty h}$ & $-0.344$ & $-0.344$ & $5.16$ & $5.16$ \\
$Al_3$ & $D_{3h}$ & $-0.582$ & $-0.582$ & $5.30$ & $5.30$ \\    
\hline
\end{tabular}
\caption{Binding energy ($\mathcal{E}_b$) and equilibrium bond length ($R_e$) for the $Al_2$ and $Al_3$ clusters with symmetry group $G$.}
\label{Table:Al:Clusters}
\end{table}

\begin{figure}[H]
\centering
\includegraphics[keepaspectratio=true,width=0.6\textwidth]{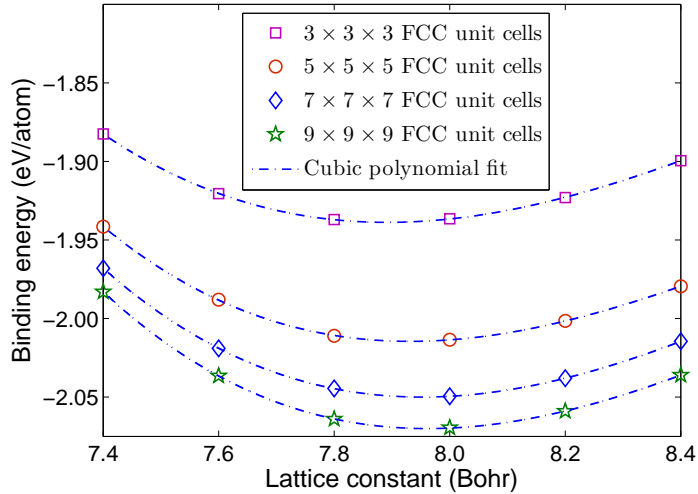}
\caption{Binding energy as a function of lattice constant for the clusters consisting of $m\times m\times m$ FCC Aluminum unit cells, where $m=3$, $5$, $7$ and $9$.}
\label{fig:BindingEnergyLatticeConstant}
\end{figure}

\begin{figure}[H]
\centering
\includegraphics[keepaspectratio=true,width=0.6\textwidth]{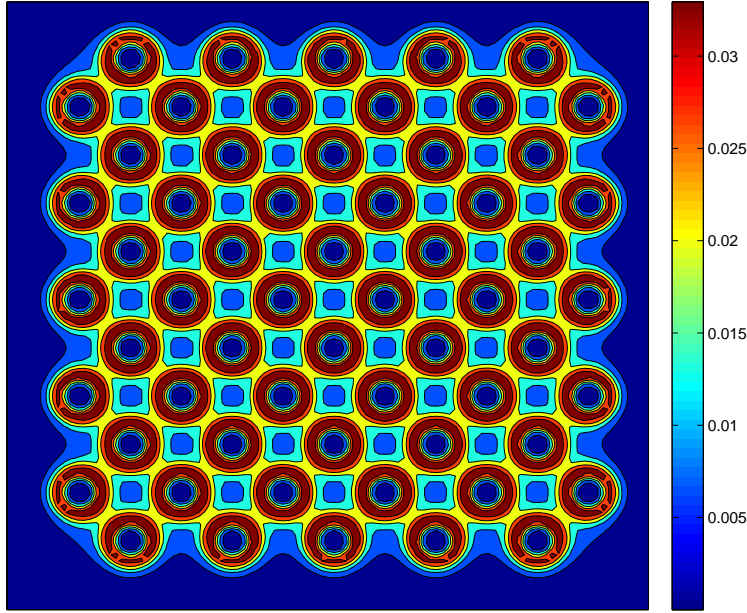}
\caption{Electron density contours on the mid-plane of a $5\times 5 \times 5$ FCC aluminum cluster with a lattice constant of $a=7.93$ Bohr.}
\label{Fig:rho_contour}
\end{figure}

Next, we study clusters consisting of $1\times1\times1$, $3\times3\times3$, $5\times5\times5$, $7\times7\times7$ and $9\times9\times9$ FCC unit cells of Aluminum with the atoms held fixed, i.e. no geometry optimization. These clusters comprise of $M=14$, $172$, $666$, $1688$ and $3430$ atoms respectively. First, we evaluate the binding energy for various lattice constants. We then fit a cubic polynomial to this data, which is utilized to calculated the equilibrium lattice constants and binding energies. The results so obtained are presented in Fig. \ref{fig:BindingEnergyLatticeConstant} and Table \ref{Table:Properties:FCC}. It is evident that the predictions of RS-FD are in very good agreement with PROFESS. In fact, the equilibrium lattice constants are identical to within $0.001$ Bohr, and the maximum difference in the binding energies is $0.001$ eV/atom. In Fig. \ref{Fig:rho_contour}, we plot the contours of electron density on the mid-plane of the $5\times5\times5$ FCC Aluminum cluster at its equilibrium lattice constant $a = 7.93$ Bohr. For this simulation, we utilize a cubical domain $\Omega$ with $L_1=L_2=L_3=64$ Bohr and sixth order accurate finite-differences with $h=0.5$ Bohr. The computational time taken on a single core of a workstation with an Intel Xeon processor ($3.4$ GHz, $12$ M L3, $6.4$ GT/s) is $10$ hours. Notably, around $95$ \% of this time was spent in the repeated solution of the Poisson equation. On the same workstation, the time taken by PROFESS with $E_{cut}=600$ eV and supercell of size $32$ Angstrom is $0.072$ hours. It is clear that a spectral scheme like plane-waves thoroughly outperforms a real-space approach like finite-differences for serial OF-DFT computations. This is due to the fact that the repeated solution of the Poisson equation --- which nearly takes up all of the time in the current real-space implementation --- is relatively inexpensive within the plane-wave method. However, in the context of high-performance computing, we anticipate real-space approaches like the one proposed here to become competitive with the plane-wave method.

\begin{table}[htbp]
\centering
\begin{tabular}{cccccc}
\hline 
FCC Aluminum & No. of atoms & $\mathcal{E}_b$ (eV/atom) & $\mathcal{E}_b$ (eV/atom) & $a_e$ (Bohr) & $a_e$ (Bohr) \\
unit cells & $(M)$ &  RS-FD & PROFESS & RS-FD & PROFESS \\
\hline
$1\times1\times1$ & $14$ & $-1.211$ & $-1.211$ & $7.73$ & $7.73$ \\
$3\times3\times3$ & $172$ & $-1.778$ & $-1.778$ & $7.89$ & $7.89$ \\
$5\times5\times5$ & $666$ & $-1.929$ & $-1.930$ & $7.93$ & $7.93$ \\
$7\times7\times7$ & $1688$ & $-2.000$ & $-2.000$ & $7.95$ & $7.95$ \\
$9\times9\times9$ & $3430$ & $-2.040$ & $-2.041$ & $7.96$ & $7.96$ \\
\hline
\end{tabular}
\caption{Binding energy ($\mathcal{E}_b$) and equlibrium lattice constant ($a_e$) for the clusters consisting of $m\times m\times m$ FCC Aluminum unit cells, where $m=1$, $3$, $5$, $7$ and $9$.}
\label{Table:Properties:FCC}
\end{table} 

\begin{figure}[H]
\centering
\includegraphics[keepaspectratio=true,width=0.48\textwidth]{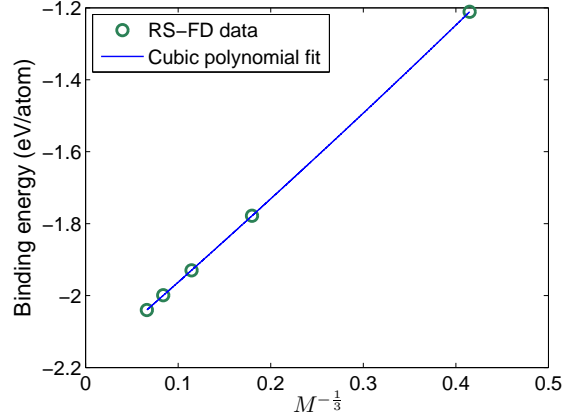}
\caption{Binding energy ($\mathcal{E}_b$) as a function of $M^{-\frac{1}{3}}$ for the clusters consisting of $m\times m\times m$ FCC Aluminum unit cells, where $m=1$, $3$, $5$, $7$ and $9$. $M$ represents the number of atoms. }
\label{Fig:scalingCohesiveEnergy}
\end{figure}

Next, we use the results presented in Table \ref{Table:Properties:FCC} to extract the bulk cohesive energy for Aluminum. We use the scaling \cite{Ahlrichs1999,Gavini2007}
\begin{equation} \label{Eqn:Scaling:CohesiveEnergy}
\mathcal{E}_b = \mathcal{E}_{coh} + a_{surf} M^{-1/3} + a_{edge} M^{-2/3} + a_{corner} M^{-1} \,
\end{equation}
where $\mathcal{E}_b$ is the binding energy per atom and $\mathcal{E}_{coh}$ is the bulk cohesive energy. Further, $a_{surf}$, $a_{edge}$ and $a_{corner}$ account for the effects of surfaces, edges and corners respectively. In Fig. \ref{Fig:scalingCohesiveEnergy}, we plot both the computed binding energy $\mathcal{E}_b$ as a function of $M^{-1/3}$ and its curve fit using a cubic polynomial. From this fit, we obtain $\mathcal{E}_{coh} = -2.189$ eV/atom, which is in very good agreement with the bulk cohesive energy of $\mathcal{E}_{coh} = -2.190$ eV/atom obtained by PROFESS for a FCC Aluminum crystal. These results have been summarized in Table \ref{Table:Scaling:PROFESS}. It is worth noting that the predicted cohesive energy for Aluminum is substantially different from both DFT \cite{goodwin1990pseudopotential} as well as experiments \cite{haberland1994clusters}, which is a limitation of the TFW kinetic energy functional. Incorporation of kernel energies, and in particular the WGC functional \cite{Wang1998,Wang1999}, are expected to provide a more accurate description. However, this is expected to come at significant additional computational cost.

\begin{table}[H]
\centering
\begin{tabular}{ccc}
\hline 
Property & RS-FD (Scaling relation) & PROFESS \\
\hline
Bulk cohesive energy (eV/atom) & $-2.189$ & $-2.190$ \\
\hline
\end{tabular}
\caption{Comparison of bulk cohesive energy obtained by RS-FD via the scaling relation (Eqn. \ref{Eqn:Scaling:CohesiveEnergy}) and a crystal calculation by PROFESS for FCC Aluminum.}
\label{Table:Scaling:PROFESS}
\end{table} 

Finally, we consider a Aluminum cluster consisting of $102690$ atoms arranged as $29\times29\times29$ FCC unit cells with lattice constant of $a=8.0$ Bohr. For this simulation, we choose $h=0.5$ Bohr, sixth order accurate finite differences and a cubic domain $\Omega$ with $L_1=L_2=L_3=128$ Bohr. This translates to roughly $135$ million finite-difference nodes in the simulation domain. We utilize $10$ nodes of a computer cluster wherein each node has the following configuration: Altus 1804i Server - 4P Interlagos Node, Quad AMD Opteron 6276, 16C, 2.3 GHz, 128GB, DDR3-1333 ECC, 80GB SSD, MLC, 2.5" HCA, Mellanox ConnectX 2, 1-port QSFP, QDR, memfree, CentOS, Version 5, and connected through InfiniBand cable. The number of matrix-vector products required for convergence is approximately $2.39$ million and wall-clock time is around $340$ hours. The binding energy so obtained is reported in Table \ref{Table:FCC29}. This example demonstrates that with the aid of high performance computing, the proposed approach represents a viable choice of performing large-scale OF-DFT calculations.

\begin{table}[H]
\centering
\begin{tabular}{cccc}
\hline 
 Cluster & $M$ & $a$ (Bohr) & $\mathcal{E}_b$ (eV/atom) \\
\hline
 $29\times29\times29$ FCC Aluminum & $102690$ & $8.0$ & $-2.143$ \\
\hline
\end{tabular}
\caption{Binding energy ($\mathcal{E}_b$) obtained by RS-FD for a cluster consisting of $29\times29\times29$ FCC unit cells of Aluminum with lattice constant $a=8.0$ Bohr.}
\label{Table:FCC29}
\end{table} 

\section{Concluding Remarks} \label{Section:Conclusions}
We have developed an Augmented Lagrangian formulation of orbital-free Density Functional Theory (OF-DFT). In particular, we have employed the Augmented Lagrangian technique to convert OF-DFT's constrained minimization problem into a sequence of unconstrained minimization problems, thereby making it amenable to powerful unconstrained optimization techniques like conjugate gradients. We have also developed a real-space, non-periodic, parallel implementation of this formulation in the framework of higher-order finite-differences and the conjugate gradient method. Using this implementation, we have demonstrated that the proposed Augmented Lagrangian approach is highly competitive with the penalty and Lagrange multiplier methods that have previously been utilized for OF-DFT. Additionally, we have shown that the higher rates of convergence with spatial discretization that are typically required for performing efficient OF-DFT calculations can be achieved with higher-order finite-differences. Finally, we have verified through selected examples that the proposed approach is not only robust and efficient, but also capable of obtaining the chemical accuracies desired in electronic structure calculations. 

The current Augmented Lagrangian formulation and implementation of OF-DFT is restricted to the TFW kineric energy functional and non-periodic boundary conditions. In order to obtain accurate material properties, nonlocal kinetic energy functionals like WGC need to be incorporated in conjunction with the ability to handle periodic boundary conditions. This is the subject of current work by the authors. Another limitation of the current implementation is that it does not achieve perfect linear-scaling in practice. Overcoming this shortcoming requires the development of effective real-space preconditioners, a worthy subject for future research. 

\section*{Acknowledgements}
Phanish Suryanarayana gratefully acknowledges the support of National Science Foundation under Grant Number $1333500$.

\appendix 
\section{Electrostatic correction for overlapping charge density of nuclei} \label{Appendix:Correct:RepulsiveEnergy}
Within the local reformulation of the electrostatics presented in Section \ref{Sec:OFDFT}, the repulsive energy of the nuclei can be expressed as \cite{Phanish2012}
\begin{equation} \label{Eqn:EZZ:rf}
\mathcal{E}_{\rm zz}(\bR) = \frac{1}{2} \int \int \frac{b(\bx,\bR) b(\bx',\bR)}{|\bx-\bx'|} \, \mathrm{d\bx} \, \mathrm{d\bx'} - \frac{1}{2}\sum_{I=1}^{M} \int b_I(\bx,\bR_I) V_I(\bx,\bR_I) \, \mathrm{d\bx} \,,
\end{equation}
where the second term accounts for the self energy of the nuclei. Using Eqn. \ref{Eqn:b:Pseudopotential}, we arrive at
\begin{eqnarray}
\mathcal{E}_{\rm zz}(\bR) & = & \frac{1}{2} \sum_{I=1}^M \sum_{J=1}^{M} \int b_I(\bx,\bR_I) V_J(\bx,\bR_J) \, \mathrm{d\bx} - \frac{1}{2}\sum_{I=1}^{M} \int b_I(\bx,\bR_I) V_I(\bx,\bR_I) \, \mathrm{d\bx} \nonumber \\
& = & \frac{1}{2} \sum_{I=1}^M \sum_{\begin{subarray}{l} J=1 \\J \neq I \end{subarray}}^{M} \int b_I(\bx,\bR_I) V_J(\bx,\bR_J) \, \mathrm{d\bx} \,,
\end{eqnarray}
If the charge density of the nuclei do not overlap, the above expression can be rewritten as
\begin{eqnarray}
\mathcal{E}_{\rm zz}(\bR) & = &  \frac{1}{2} \sum_{I=1}^M \sum_{\begin{subarray}{l} J=1 \\J \neq I \end{subarray}}^{M} Z_J \int \frac{b_I(\bx,\bR_I)}{|\bx-\bR_J|} \, \mathrm{d\bx}  = \frac{1}{2} \sum_{I=1}^M \sum_{\begin{subarray}{l} J=1 \\J \neq I \end{subarray}}^{M} Z_J V_I(\bR_J,\bR_I)\nonumber \\
& = & \frac{1}{2} \sum_{I=1}^{M} \sum_{\begin{subarray}{l} J=1 \\J \neq I \end{subarray}}^{M} \frac{Z_{I} Z_{J}}{|\bR_{I}-\bR_{J}|} \,,
\end{eqnarray}
which is exactly repulsive energy given in Eqn. \ref{Eqn:EZZ}. However, there is a possibility that the charge density of the nuclei overlap when using the pseudopotential approximation, particularly when the cutoff radius $r_c$ is relatively large. In this situation, the repulsive energy and the forces on the nuclei will be inaccurately calculated. We present a technique below to account these errors. 

We start by generating spherically symmetric and compactly supported `reference' charge density of the nuclei, which we refer to as $\tilde{b}_J(\bx,\bR_J)$. In particular, we choose non-overlapping $\tilde{b}_J(\bx,\bR_J)$ which satisfy the relation
\begin{equation}
\int \tilde{b}_J (\bx,\bR_J) \, \mathrm{d\bx} = Z_J .
\end{equation}
In this setting, the correction to the repulsive energy and therefore the total energy can be expressed as
\begin{eqnarray} \label{Eqn:RepulsiveCorrection}
\mathcal{E}_c^*(\bR) & = & \frac{1}{2} \int \int \frac{\tilde{b}(\bx,\bR) \tilde{b}(\bx',\bR)}{|\bx-\bx'|} \, \mathrm{d\bx} \, \mathrm{d\bx'} - \frac{1}{2} \int \int \frac{b(\bx,\bR) b(\bx',\bR)}{|\bx-\bx'|} \, \mathrm{d\bx} \, \mathrm{d\bx'}  \nonumber \\
& & - \frac{1}{2}\sum_{J=1}^{M} \int \tilde{b}_J(\bx,\bR_J) \tilde{V}_J(\bx,\bR_J) \, \mathrm{d\bx}  + \frac{1}{2}\sum_{J=1}^{M} \int b_J(\bx,\bR_J) V_J(\bx,\bR_J) \, \mathrm{d\bx} \,,
\end{eqnarray}
where 
\begin{equation}
\tilde{b}(\bx,\bR) = \sum_{J=1}^M \tilde{b}_J(\bx,\bR_J) \,,
\end{equation}
A direct computation of this energy correction will scale quadratically with respect to the number of atoms. In order to ensure linear-scaling, we rewrite Eqn. \ref{Eqn:RepulsiveCorrection} as
\begin{eqnarray} \label{Eqn:RepulsiveCorrection2}
\mathcal{E}_c^*(\bR) & = & \frac{1}{2} \int \left( \tilde{b}(\bx,\bR) + b(\bx,\bR) \right) V_c(\bx,\bR) \, \mathrm{d\bx} + \frac{1}{2}\sum_{J=1}^{M} \int b_J(\bx,\bR_J) V_J(\bx,\bR_J) \, \mathrm{d\bx} \nonumber \\
& & - \frac{1}{2}\sum_{J=1}^{M} \int \tilde{b}_J(\bx,\bR_J) \tilde{V}_J(\bx,\bR_J) \, \mathrm{d\bx} \,,
\end{eqnarray}
where $V_c(\bx,\bR)$ is the solution to the Poisson equation 
\begin{equation}
\frac{-1}{4\pi} \nabla^2 V_c(\bx,\bR) = \tilde{b}(\bx,\bR) - b(\bx,\bR).
\end{equation} 

The correction to the forces on the nuclei can be expressed as 
\begin{eqnarray}
{\bf f}_J^c & = & -\frac{\partial \mathcal{E}_c^*(\bR)}{\partial \bR_J} \\
& = & -\int \frac{\partial \tilde{b}_J(\bx,\bR_J)}{\partial \bR_J}\left(\tilde{V}(\bx,\bR)- \tilde{V}_J(\bx,\bR_J)\right) \, \mathrm{d\bx} + \int \frac{\partial b_J(\bx,\bR_J)}{\partial \bR_J}\left(V(\bx,\bR)- V_J(\bx,\bR_J)\right) \, \mathrm{d\bx} \nonumber \\
& = & \int \nabla \tilde{b}_J(\bx,\bR_J) \left(\tilde{V}(\bx,\bR)- \tilde{V}_J(\bx,\bR_J)\right) \, \mathrm{d\bx} - \int \nabla \tilde{b}_J(\bx,\bR_J)\left(V(\bx,\bR)- V_J(\bx,\bR_J)\right) \, \mathrm{d\bx} \nonumber
\end{eqnarray}
where $\tilde{V}(\bx,\bR)$ and $V(\bx,\bR)$ can be evaluated by solving the Poisson equations
\begin{equation}
\frac{-1}{4 \pi} \nabla^2 \tilde{V}(\bx,\bR) = \tilde{b}(\bx,\bR) \,, \quad  \frac{-1}{4 \pi} \nabla^2 V(\bx,\bR) =  b(\bx,\bR)  .
\end{equation}
It is important to note that even with these corrections to the energy and forces, the overall OF-DFT formulation maintains its linear-scaling with respect to the number of atoms.

\bibliographystyle{ReferenceStyle}
\bibliography{AugLagOFDFT}

\end{document}